\documentclass[11pt]{article}
\usepackage{amsmath,amssymb}
\usepackage{epsfig}
\begin{document}
%\begin{titlepage}
\begin{flushright}   SU-4252-756
	     \\
\end{flushright}	 
\begin{flushright}   May 2002
             \\
\end{flushright}   
\begin{center}
   \vskip 3em
{\LARGE THE STAR PRODUCT ON THE FUZZY}
\vskip 1em
{\LARGE SUPERSPHERE}
  \vskip 3em
{\large A. P. Balachandran, S. K\"{u}rk\c{c}\"{u}o\v{g}lu and E. Rojas \footnote{E-mails: bal@phy.syr.edu,
skurkcuo@phy.syr.edu, erojas@phy.syr.edu} 
\\[3em]}
\em{
Department of Physics, Syracuse University,
Syracuse, NY 13244-1130, USA}
\end{center}
\vskip 2em
\begin{abstract}
The fuzzy supersphere $S_F^{(2,2)}$ is a finite-dimensional matrix approximation
to the supersphere $S^{(2,2)}$ incorporating supersymmetry exactly. Here
the $\star$-product of functions on $S_F^{(2,2)}$ is
obtained by utilizing the $OSp(2,1)$ coherent states. We check its
graded commutative limit to $S^{(2,2)}$ and extend it to fuzzy versions of sections of
bundles using the methods of \cite{private}. A brief discussion of the
geometric structure of our $\star$-product completes our work.
\end{abstract} 
%PACS: 04.20.Fy, 31.15.Pf, 11.10.Ef
%\end{titlepage}
\newpage
\setcounter{footnote}{0}

\section{Introduction}

Studies of field theories on non-commutative (fuzzy) 
manifolds have started to produce many novel and encouraging
results in the last few years. In these models
one takes compact manifolds which are usually co-adjoint
orbits of Lie groups and discretises them by quantization
\cite{madore}.

Upon quantization the discretized manifold exhibits a 
cell-like structure with the number of cells being finite. 
In this new non-commutative geometry without points, all 
symmetries of a field theory are generally preserved. With these 
properties, this recently divised technique could serve
as a non-perturbative regulator for field theories \cite{ydri,
regulator}.

Among these manifolds, the fuzzy $2$-sphere $S_F ^2$ has 
been extensively studied. It has become evident that 
field theories on $S_F ^2$ avoid fermion doubling as
well as permit reformulation of sigma models and extended
objects such as monopoles, instantons, etc \cite{monopole, fermion, bott}.
 
An important ingredient in understanding the geometrical 
structure underlying these fuzzy models and their 
continuum limit is the associative $\star$-product of 
functions. Recently, explicit expressions for
$\star$-product of functions on $S_F ^2$, $CP_F ^{N}$
(the fuzzy $CP_N$) and the fuzzy complex Grassmannian
spaces have appeared in the literature \cite{chiral, BDBJ, grassmann}.
One way that has been followed in these studies was first to
introduce generalized Perelomov-like coherent states
\cite{perelomov} and map an operator to a function on a fuzzy
manifold by identifying its diagonal matrix elements with values
of the function on the fuzzy manifold. It is a theorem that
diagonal coherent state elements of operators completely
determine that operator. An associative $\star$-product of two
such functions is therefrom introduced straightforwardly
and their properties are studied in detail. A finite series
expansion of this product is obtained by the authors in terms of
the derivatives of the functions involved and a projection operator
enclosing the differential geometric structure of the manifold.

In this article we investigate the construction of
an associative $\star$-product of functions
on the fuzzy supersphere $S_F ^{(2,2)}$. The latter have been
studied in the articles by Grosse et al.
\cite{GKP1, fuzzyS}. Our formulation of the problem will be
based on \cite{GKP1} for the properties of the fuzzy
$S_F ^{(2,2)}$ and the work \cite{BDBJ} for the introduction
of the $\star$-product. Our construction of the coherent
states on the supergroup $OSp(2,1)$ will rely on the use of
annihilation-creation operators, however their equivalence
to the $OSp(2,1)$ supergroup-induced coherent states will be explicitly
shown. We also extend our results to obtain $\star$-products
on ``sections of bundles'' on $S_F ^{(2,2)}$ using the methods of
\cite{private}.

Our text is organized as follows. In Section 2, to fix notation 
and conventions and to be self contained we briefly review the 
representation theory and basic properties of Lie 
superalgebras $osp(2,1)$ and $osp(2,2)$ and the
corresponding supergroups $OSp(2,1)$ and $OSp(2,2)$, 
which underlie the constrution of $S_F ^{(2,2)}$. In Section 3, we take
on the task of constructing the supercoherent states which will be
used to induce the definition of the $\star$-product in a later
section. Section 4 briefly summarizes the definition and properties of
the usual and fuzzy superspheres $S^{(2,2)}$ and $S_F ^{(2,2)}$
respectively from a group theoretic point of view. In Section 5,
we introduce the $\star$-product
on $S_F ^{(2,2)}$ and compute it by utilizing the properties of
supercoherent states of Section 3. The product and its
properties are discussed in detail. This is followed by
a discussion of (fuzzy) sections
of bundles and the form of the $\star$-product for their
elements. Section 6 includes remarks on the differential
geometric structure underlying the $\star$-product. Some observations and
discussion of further directions we are planning to explore in
forthcoming studies conclude our work.

\section{$osp(2,1)$ and $osp(2,2)$ Superalgebras and
their Associated Supergroups}

Here we review some of the basic facts regarding the
Lie superalgebras $osp(2,1)$ and $osp(2,2)$ and their
associated supergroups $OSp(2,1)$ and $OSp(2,2)$.
For detailed discussions, the reader is refered to the references
\cite{Dewitt, cornwell, pais, nahm1, nahm2}.

\subsection{ $osp(2,1)$ and $osp(2,2)$ Superalgebras
and their Representations}

Representations and properties of the Lie superalgebras
have well-but not widely known features that we would like
to briefly review for our purposes. As for any graded Lie algebra,
$osp(2,1)$ and $osp(2,2)$ have even and odd parts. The even
part of $osp(2,1)$ is the Lie algebra $su(2)$ with its usual
generators $\Lambda_i$ ($i,j= 1,2,3$). Its odd part is built up
of $su(2)$ spinors $\Lambda_\alpha$ ($\alpha,\beta= 4,5$). They
fulfill further properties to be explained below.
The graded commutation relations are \cite{pais, nahm2}
\begin{eqnarray}
\lbrack \Lambda_i ,\Lambda_j \rbrack &=& i \epsilon_{ijk} \Lambda_k \,, \\
\lbrack\Lambda_i , \Lambda_\alpha \rbrack &=& \frac{1}{2}
(\sigma_i)_{ \beta \alpha} \Lambda_\beta  \,, \\
\{ \Lambda_\alpha , \Lambda_\beta \} &=& \frac{1}{2}
(C \sigma_i)_{\alpha \beta} \Lambda_i \,, 
\end{eqnarray}
where $\sigma_i$ are the Pauli matrices and 
$C_{\alpha \beta} = - C_{\beta \alpha}$ is the Levi-Civita symbol with $C_{45} = 1$.
(We use the indices 4,5 for their rows and columns.)

In the graded Lie algebras of our interest, the usual adjoint
(or star) operation $\dagger$ on Lie algebras is replaced by the grade adjoint
(or grade star) operation $\ddagger$ \cite{nahm1}. First, we note that
the grade adjoint of an even (odd) element is even (odd). Next, one has
$A ^{\ddagger \ddagger} = (-1)^{|A|} A$ for an even or odd (that is homogeneous)
element $A$ of degree $|A|$ $(mod \,2)$, or equally well,
integer $(mod \,2)$. (So, depending on $|A|$, $|A|$ itself can be taken $0$ or $1$.) Thus, it is the usual
$\dagger$ on the even part, while on odd elements $A$, it squares to $-1$.
Furthermore, ${\lbrack A, B \}}^\ddagger
= (-1)^{|A||B|} \lbrack B^\ddagger, A^\ddagger \}$ for homogeneous elements
$A, B$, $\lbrack A, B \}$ denoting the graded Lie bracket.

Henceforth we will denote the degree of $a$ (which may be a super Lie algebra element, a linear operator
or an index) by $|a| (mod \,2)$, $|a|$ denoting any integer in its equivalance class
$< |a| + 2n \,: n \in {\mathbb Z}>$.

Following \cite{nahm1, nahm2} we remark
that any element of the $osp(2,1)$ (and $osp(2,2)$) graded Lie algebras has to fulfill certain ``reality''
properties implemented by $\ddagger$. For the generators of $osp(2,1)$
these are given by
\begin{equation}
\Lambda_i^{\ddagger} = \Lambda_i^{\dagger} = \Lambda_i,
\quad \quad
\Lambda_\alpha ^{\ddagger} = \sum_{\beta = 4,5}C_{\alpha \beta} \Lambda_\beta \, \quad \alpha = 4,5 \,.
\label{eq:six}
\end{equation}

In a (grade star) representation of a graded Lie algebra on a graded vector space $V$, let
$V = V_0 \oplus V_1$ where $V_0$ and $V_1$ are even and odd subspaces \cite{cornwell}.
$V_0$ and $V_1$ are invariant under the even elements of the graded
Lie algebra while its odd elements map one to the other. Let us also
assume that $V$ is endowed with the inner product $<u|v>$ for all
$u, v \in V$. Now if $L$ is a linear operator acting on $V$ then the grade
adjoint of $L$ is defined by
\begin{equation}
<L^\ddagger \,u | v> = (-1)^{|u| \,|L|}\, <u|L \,v> \,.
\end{equation}
In a basis adapted to the above decomposition of $V$, $L$ has the matrix representation
\begin{equation}
M_L = \left(
\begin{array}{cc}
\alpha_1 & \alpha_2 \\
\alpha_3 & \alpha_4
\end{array}
\right) = M_0 + M_1 \quad \quad 
M_0 = \left(
\begin{array}{cc}
\alpha_1 & 0 \\
0 & \alpha_4
\end{array}
\right) \,, \quad \quad
M_1 = \left(
\begin{array}{cc}
0 & \alpha_2 \\
\alpha_3 & 0
\end{array}
\right)
\end{equation}
where $M_0$ and $M_1$ are the even and odd parts of $M_L$.
The formula for $\ddagger$ is then
\begin{equation}
M_L^{\ddagger} = \left(
\begin{array}{cc}
\alpha_1^{\dagger} & \alpha_3^{\dagger} \\
-\alpha_2^{\dagger} & \alpha_4^{\dagger}
\end{array}
\right) \,,
\end{equation}
$\alpha_i^{\dagger}$ being matrix adjoint of $\alpha_i$. We note
that the supertrace $str$ of $M_L$ is:
\begin{equation}
strM_L = Tr \alpha_1 - Tr\alpha_4 \,.
\end{equation} 
   
The irreducible representations of $osp(2,1)$ 
\cite{ GKP1, pais, nahm2, klimcik} are characterized
by an integer or half-integer non-negative quantum  number $J_{osp(2,1)}$
called superspin. From the point of view of the irreducible 
representations of $su(2)$, the superspin $J_{osp(2,1)}$ representation
has the decomposition
\begin{equation}
J_{osp(2,1)} = J_{su(2)} \oplus \left(J - \frac{1}{2}
\right)_{su(2)}\,,
\end{equation}
where $J_{su(2)}$ is the $su(2)$ representation for angular momentum $J_{su(2)}$.
In particular the fundamental and adjoint 
representations of $osp(2,1)$ correspond to $J_{osp(2,1)}=1/2$ and $J_{osp(2,1)}=1$ respectively,
being 3 and 5 dimensional. The quadratic Casimir operator
is given by
\begin{equation}
K_2 = \Lambda_i \Lambda_i + C_{\alpha \beta} \Lambda_\alpha 
\Lambda_\beta.
\end{equation}
It has the eigenvalues $J(J + 1/2)$.

The $osp(2,2)$ superalgebra \cite{GKP1, nahm2, klimcik} can 
be defined by introducing 
an even generator $\Lambda_8$ commuting with the $\Lambda_i$
and odd generators $\Lambda_\alpha$  with $\alpha = 6,7$ in
addition to the already existing ones for $osp(2,1)$. The graded 
commutation relations for $osp(2,2)$ are then
\begin{eqnarray}
\lbrack \Lambda_i ,\Lambda_j \rbrack &=& i \epsilon_{ijk} \Lambda_k \\
\lbrack\Lambda_i , \Lambda_\alpha \rbrack &=& \frac{1}{2} 
(\tilde{\sigma}_i)_{ \beta \alpha} \Lambda_\beta \\
\lbrack \Lambda_i ,\Lambda_8 \rbrack &=& 0 \\
\lbrack \Lambda_8 ,\Lambda_\alpha \rbrack &=& 
\tilde{\varepsilon}_{\alpha \beta} \Lambda_\beta \\
\{ \Lambda_\alpha ,\Lambda_\beta \} &=& \frac{1}{2} 
(\tilde{C} \tilde{\sigma_i})_{\alpha \beta} \Lambda_i 
+ \frac{1}{4} (\tilde{\varepsilon}\tilde{C})_{\alpha \beta} \Lambda_8
\end{eqnarray}
where $i,j= 1,2,3$ and $\alpha, \beta= 4,5,6,7$. In above
we have used the matrices
\begin{equation}
\tilde{\sigma}_i = \left(
\begin{array}{cc}
\sigma_i & 0 \\
0 & \sigma_i
\end{array}
\right) 
\quad \quad 
\tilde{C} = \left(
\begin{array}{cc}
C & 0 \\
0 & -C
\end{array} 
\right)
\quad \quad
\tilde{\varepsilon} = \left(
\begin{array}{cc}
0 & I_{2\times 2} \\
 I_{2\times 2}& 0
\end{array} 
\right)\,.
\label{eq:matrices}
\end{equation}
Their matrix elements are indexed by $4,\ldots ,7$.

In addition to (\ref{eq:six}), the new generators satisfy
the ``reality" conditions
\begin{equation}
\Lambda_\alpha ^{\ddagger} = - \sum_{\beta = 6,7} \tilde{C}_{\alpha \beta}
\Lambda_\beta \,,\quad \alpha = 6,7 \,, \quad \quad
\Lambda_8^{\ddagger} = \Lambda_8^{\dagger} = \Lambda_8 \,.
\label{eq:reality1}
\end{equation}
So we can write the $osp(2,2)$ reality conditions for all $\alpha$ as 
$\Lambda_\alpha ^\ddagger = \tilde{C}_{\alpha \beta} \Lambda_\beta$.
 
Irreducible representations of $osp(2,2)$ fall into two categories,
namely the typical and non-typical ones \cite{GKP1, klimcik}.
Typical ones are 
reducible with respect to the $osp(2,1)$ superalgebra
(except for the trivial representation) whereas 
non-typical ones are irreducible. Typical representations 
are labeled by an integer or half integer non-negative number 
$J_{osp(2,2)}$, called $osp(2,2)$ superspin. These have the 
$osp(2,1)$ content $J_{osp(2,2)}= J_{osp(2,1)} \oplus 
(J - 1/2)_{osp(2,1)}$ for $J_{osp(2,2)} \geq 1/2$ while $(0)_{osp(2,2)}=
(0)_{osp(2,1)}$. Hence
\begin{equation}
J_{osp(2,2)} = \left\{
\begin{array}{ll}
J_{su(2)} \oplus \left( J - \frac{1}{2}\right)_{su(2)}
\oplus \left( J - \frac{1}{2}\right)_{su(2)} \oplus (J - 1)_{su(2)} \,,
& J_{osp(2,2)} \geq 1  \,; \\
(\frac{1}{2})_{su(2)} + (0)_{su(2)} + (0)_{su(2)} \,,
& J_{osp(2,2)} = \frac{1}{2} \,.
\end{array} 
\right.
\end{equation}
$osp(2,2)$ has the quadratic Casimir operator
\begin{eqnarray}
K_2 ' &=& \Lambda_i \Lambda_i + \tilde{C}_{\alpha \beta}
\Lambda_\alpha \Lambda_\beta - \frac{1}{4} \Lambda_8 ^2 
\nonumber \\
&=& K_2 - (\sum_{\alpha, \beta = 6,7} -\tilde{C}_{\alpha \beta}
\Lambda_\alpha \Lambda_\beta + \frac{1}{4} \Lambda_8 ^2) \,.
\end{eqnarray} 
Note that since all the generators of $osp(2,1)$
commute with $K_2 '$ and $K_2$, they also commute with
\begin{equation}
V:= -\sum_{\alpha, \beta = 6,7} \tilde{C}_{\alpha \beta}
\Lambda_\alpha \Lambda_\beta + \frac{1}{4} \Lambda_8 ^2 \,.
\end{equation}

As regards the non-typical representations of $osp(2,2)$ associated with $J_{osp(2,1)}$, 
we note that the substitutions
\begin{equation}
\Lambda_i \rightarrow \Lambda_i, \quad \Lambda_\alpha \rightarrow \Lambda_\alpha, \quad 
\alpha = 4,5; \quad \Lambda_\alpha \rightarrow - \Lambda_\alpha, \quad \alpha = 6,7; \quad
\Lambda_8 \rightarrow - \Lambda_8 \,
\end{equation}
define an automorphism of $osp(2,2)$. This automorphism changes a non-typical
irreducible representation into an inequivalent one (except for the trivial representation with $J = 0$),
while preserving the reality conditions given
in (\ref{eq:six}) and (\ref{eq:reality1}) \cite{nahm2}.
We discriminate between these
two representations  associated with $J_{osp(2,1)}$ as follows: For $J>0$, 
$\hat{J}_{osp(2,2) + }$ will denote the representation in which the eigenvalue of
the representative of $\Lambda_8$ on vectors with angular momentum $J$ is positive
and $\hat{J}_{osp(2,2) - }$ will denote its partner where this eigenvalue is negative.
(This eigenvalue is zero only in the trivial representation with $J=0$.) 
In this paper we concentrate on  $\hat{J}_{osp(2,2) + }$.
The results for  $\hat{J}_{osp(2,2) - }$ are similar and will be occasionally indicated.
     
Below we list some of the well-known results and standard notations that are used 
throughout the text \cite{pais, nahm2, chaichian}.
The fundamental representation of $osp(2,2)$ is non-typical and
we concentrate on the one given by
$\hat{J}_{osp(2,2) + } = \hat{(\frac{1}{2})}_{osp(2,2) + }$.
%(We denote the non-typicalrepresentations of $osp(2,2)$ associated with
%$J_{osp(2,1)}$ by $\hat{J}_{osp(2,2)}$).   
It is generated by the$(3 \times 3)$ supertraceless matrices $\Lambda_a ^{(\frac{1}{2})}$
satisfying the ``reality'' conditions of (\ref{eq:six}) and (\ref{eq:reality1}):
\begin{eqnarray}
\Lambda_i ^{(\frac{1}{2})} = \frac{1}{2}\left(
\begin{array}{cc}
 \sigma_i & \,0 \\
\,0 & 0
\end{array}
\right)
\quad \quad  
\Lambda_4  ^{(\frac{1}{2})} & = &\frac{1}{2} \left(
\begin{array}{cc}
0 & \xi \\
\eta^T & 0
\end{array} 
\right)
\quad \quad
\Lambda_5  ^{(\frac{1}{2})} =  \frac{1}{2} \left(
\begin{array}{cc}
0 & \eta \\
-\xi^T & 0
\end{array} 
\right) \nonumber \\
\Lambda_6 ^{(\frac{1}{2})} = \frac{1}{2} \left(
\begin{array}{cc}
0 & -\xi \\
\eta^T & 0
\end{array} 
\right)
\quad \quad
\Lambda_7 ^{(\frac{1}{2})} & = & \frac{1}{2} \left(
\begin{array}{cc}
0 & -\eta \\
-\xi^T & 0
\end{array}
\right)
\quad \quad
\Lambda_8 ^{(\frac{1}{2})} = \left(
\begin{array}{cc}
I_{2\times 2} & 0 \\
\,0 & 2
\end{array}
\right) 
\label{eq:nontyp}
\end{eqnarray}
where
\begin{equation}
\xi = \left(
\begin{array}{c}
-1 \\
0
\end{array}
\right)
\quad \quad {\mbox{and}} \quad \quad
\eta = \left(
\begin{array}{c}
0 \\
-1
\end{array}
\right) \,.
\end{equation}
These generators satisfy 
\begin{equation}
\Lambda_a ^{(\frac{1}{2})} \Lambda_b ^{(\frac{1}{2})}
= S_{ab}{\bf 1} + \frac{1}{2}\, (d_{abc} + 
i f_{abc} )\,\Lambda_c  ^{(\frac{1}{2})} \quad \quad (a,b,c = 1,2,\dots 8)
\end{equation}
with
\begin{eqnarray}
S_{ab} &=& str(\Lambda_a ^{(\frac{1}{2})} \Lambda_b ^{(\frac{1}{2})}) \,, \nonumber \\
f_{abc} &=& str(- i \lbrack \Lambda_a ^{(\frac{1}{2})} , 
\Lambda_b ^{(\frac{1}{2})} \} \Lambda_c ^{(\frac{1}{2})}) \,, \\
d_{abc} &=& str( \{ \Lambda_a ^{(\frac{1}{2})} , 
\Lambda_b ^{(\frac{1}{2})} \rbrack \Lambda_c ^{(\frac{1}{2})}) \,. \nonumber 
\end{eqnarray}
Here $a = i = 1,2,3$, and $a =8$ label the even generators and
$a = \alpha = 4,5,6,7$ label the odd generators.
Also here and what follows, $[A,B\} , \{A,B]$ denote the graded commutator and
anticommutator respectively: If $A$ and $B$ are homogeneous elements, $[A,B\} = AB - (-1)^{|A||B|} BA,
\{A,B] = AB + (-1)^{|A||B|} BA$.
 
$S_{ab}$ defines the invariant metric of the
super Lie algebra $osp(2,2)$. In their block diagonal
form, $S$ and its inverse read
\begin{equation}
S = \left(
\begin{array}{ccc}
\frac{1}{2}\,I & & \\
 & -\frac{1}{2}\,\tilde{C} & \\
 & & -2
 \end{array}
\right) _{8 \times 8}
%\quad {\mbox{and}}
\quad \quad
S ^{-1} = \left(
\begin{array}{ccc}
2\,I & & \\
 & 2\,\tilde{C} & \\
 & & -\frac{1}{2}
\end{array}
\right) _{8 \times 8} \,.
\end{equation} 
The explicit values of the structure constants $f_{abc}$ can be 
read from (\ref{eq:matrices}), since $\lbrack \Lambda_a, \Lambda_b \} = i f_{abc}
\Lambda_c$. Those of $d_{abc}$ are as follows \footnote{The tensor $d_{abc}$
given explicitly in (\ref{eq:dtensor}) for $\hat{J}_{osp(2,2) + }$ becomes $- d_{abc}$
for $\hat{J}_{osp(2,2) - }$.}:
\begin{eqnarray}
&&d_{ij8} = - \frac{1}{2} \delta_{ij} \,, \quad \quad d_{\alpha \beta 8} = \frac
{3}{4} \tilde{C}_{\alpha \beta} \,, \quad \quad 
d_{\alpha 8 \beta} = 3 
\delta_{\alpha \beta} \,, \quad d_{i8j} = 
2 \delta_{ij} \,, \nonumber \\
&&d_{\alpha \beta i}  = - \frac{1}{2}(\tilde{\varepsilon} \tilde{C}
\tilde{\sigma}_i)_{\alpha\beta} \,, \quad \quad  d_{i \alpha \beta} =
- \frac{1}{2}(\tilde{\varepsilon} \tilde{\sigma}_i)_{\beta \alpha} \,,
\quad \quad d_{888} = 6 \,.
\label{eq:dtensor}
\end{eqnarray}

We close this subsection with a final remark. Discussion in the subsequent
sections will involve the use of linear operators acting on the adjoint representation of $osp(2,2)$.
These are linear operators $\hat{\cal Q}$ acting on $\Lambda_a$ according to
$\hat{\cal Q} \Lambda_a = \Lambda_b {\cal Q}_{ba}$, ${\cal Q}$ being the matrix representation
of $\hat{\cal Q}$. They are graded because $\Lambda_a$'s are, and hence the linear operators on the
adjoint representation are graded.
The degree (or grade) of a matrix ${\cal Q}$ with only the nonzero entry ${\cal Q}_{ab}$ is
$(|\Lambda_a| + |\Lambda_b|) \, (mod \,2) \equiv (|a| + |b|) \,(mod \,2)$.
The grade star operation on $\hat{\cal Q}$ now follows from the sesquilinear form
\begin{equation}
(\alpha = \alpha_a \Lambda_A, \,\beta= \beta_b \Lambda_b) = \bar{\alpha}_a S_{ab} \beta_b,
\quad \quad \alpha_a, \beta_b \in {\mathbb C} \,
\end{equation}
and is given by
\begin{equation}
(\hat{\cal Q}^\ddagger \alpha, \beta) = (-1)^{|\alpha| \,|\hat{\cal Q}|} (\alpha, \hat{\cal Q} \beta) \,. 
\end{equation}
  
\subsection{Passage to Supergroups}

We recollect here the passage from these superalgebras to the 
corresponding supergroups \cite{cornwell, chaichian}.
Let $\xi_a$,($a =1,\ldots,8$) be the super coordinates. Here 
$a = i = 1,2,3$ and $a =8$ label the even and
$a = \alpha = 4,5,6,7$ label the odd coordinates.  
$ \xi_a$ satisfy the graded commutation relations mutually and 
with the algebra elements:
\begin{equation}
\lbrack \xi_a,\xi_b \} = 0 \,, \quad \quad \lbrack \xi_a,\Lambda_b \} = 0 \,. 
\end{equation}
We assume that $\xi_i^{\ddagger} = \xi_i,  \xi_8^{\ddagger} =
\xi_8$ and $\xi_\alpha^{\ddagger} = 
\tilde{C}_{\alpha \beta} \xi_\beta$. Then $\xi_a \Lambda_a$ is 
grade star even:
\begin{equation}
(\xi_a \Lambda_a)^{\ddagger} = \xi_a \Lambda_a.
\label{eq:hermit}
\end{equation}
(\ref{eq:hermit}) corresponds to the usual hermiticity property of Lie 
algebras which yields unitary representations of the group. 
The supergroup elements $g$ are $e^{i \xi_a \Lambda_a}$
and products of such elements. Note that $g^{\ddagger} = g^{-1}$.
      
\section{Coherent States}

In this section we construct the supercoherent states (SCS)
that are appropriate for our purposes. Alternative approaches 
for constructing $OSp(2,1)$ coherent states are given in the 
literature \cite{chaichian, gradechi}. The SCS constructed here 
will be used heavily in the following sections.

We start our discussion by introducing the coherent state
including the bosonic and fermionic degrees of freedom 
\cite{perelomov, Klauder}:
\begin{equation}
|\psi > \equiv |z,\theta> = \frac{e^{-1/2 \,|\psi|^2}}{|\psi|}\, 
e^{a_\alpha ^\dagger z_\alpha + b^\dagger \theta}\,|0> \,,
\label{eq:coherentstate}
\end{equation}
\begin{equation}
|\psi|^2 \equiv |z_1|^2 + |z_2|^2 + \bar{\theta} \theta \,.
\end{equation}
Here $a_\alpha$, $a_\alpha ^\dagger$ \,$(\alpha =1,2)$, $b$, $b^\dagger$ are
bosonic and fermionic annihilation and creation operators with the usual 
commutation and anticommutation relations
\begin{equation}
\lbrack a_\alpha ,a_\beta ^\dagger \rbrack = 
\delta_{\alpha \beta} \,,\quad \quad \{ b , b^\dagger \}=1 \,, 
\quad \quad \lbrack a_\alpha , b \rbrack = 0
\end{equation}
etc. $\theta$ is a Grassmann number such that $\{ \theta ,
\bar{\theta} \}=0$ and $\theta \theta = \bar{\theta} 
\bar{\theta} =0$. We have also fixed the normalization of $ |\psi >$'s:
\begin{equation}
<\psi|\psi> = \frac{1}{|\psi|^2} \,.
\end{equation}

In order to define our supercoherent states we require a 
finite-dimensional setting. For this purpose we will project 
the state in (\ref{eq:coherentstate}) to the $N$-dimensional 
Fock space. The projection operator is given by
\begin{equation}
P_N = \sum_{N = n_1+n_2+n_3} 
\frac{1}{n_1 ! \,n_2 !}\,(a_1 ^\dagger)^{n_1}(a_2 ^\dagger)^{n_2}
(b ^\dagger)^{n_3}|0><0|(b)^{n_3}(a_2)^{n_2}(a_1)^{n_1}  \,.
\end{equation}
Note that $n_3 =0 \,{\mbox{or}}\,1$ and that $P_N ^2 = P_N \,, 
P_N ^\dagger = P_N$.

Projecting $|\psi >$ with $P_N$ to the $N$-dimensional Fock space
and renormalizing the result by the factor $(<\psi|P_N |\psi>)^{1/2}$,
we get
\begin{equation}
|\psi , N> = \frac{1}{\sqrt{N!}} \,
\frac{(a_\alpha ^\dagger z_\alpha + b^\dagger \theta)^N}{(|\psi|)^N}
\, |0>\,.
\label{eq:5}
\end{equation}
This is exactly the supercoherent state we are 
looking for. It can be derived in another way. Consider the following highest weight 
state in the $J_{osp(2,1)} = 1/2$ representation of $osp(2,1)$ for which $N=1$: 
\begin{equation}
|J,J_3> = |1/2 , 1/2>\,.
\label{eq:state}
\end{equation}
This is also the highest weight state in the associated non-typical
representation $\hat{J}_{osp(2,2) +} = (\hat{\frac{1}{2}})_{osp(2,2) + }$ of $osp(2,2)$. 
Consider now the action of the $OSp(2,1)$ group on (\ref{eq:state}).
This can be realized by taking $g \in OSp(2,1)$ and ${\cal U}(g)$
as the corresponding element in the $3 \times 3$ fundamental
representation. Thus let
\begin{equation}
|g> = {\cal U}(g) |1/2 , 1/2>\,,
\label{eq:gcoherent}
\end{equation}
where $|g>$ is the super analogue of the Perelomov coherent 
state \cite{chaichian}.
Write
\begin{equation}
|1/2 , 1/2> = \Psi_1 ^\dagger |0>
\end{equation}
where
\begin{equation}
\Psi^\dagger = \left(\Psi_1 ^\dagger \,, \, \Psi_2^\dagger \,, \, \Psi_0 ^\dagger
\right) \equiv \left(a_1 ^\dagger \,, \, a_2 ^\dagger \,, \, b^\dagger \right) \,.
\end{equation} 
Let also \cite{chaichian}
\begin{equation}
{\cal D}(g) = \left(
\begin{array}{ccc}
z_1 ' & - \bar{z}_2 ' & - \chi \\
z_2 ' &  \bar{z}_1 ' & - \bar{\chi} \\
\theta ' & -\bar{\theta}' & \lambda
\end{array}
\right) \,, \quad \quad \sum_{i} |z_i '|^2 + \bar{\theta} ' \theta ' = 1 
\end{equation}
be the form of the matrix of ${\cal U}(g)$ in the basis $\{\Psi_\mu ^\dagger|0>\}$ $(\mu = 1,2,0)$. 
Then
\begin{eqnarray}
|g> &=& \Psi_\mu ^\dagger \,({\cal D}(g))_{\mu 1}|0>
\nonumber \\
&=& (a_\alpha ^\dagger z_\alpha ' + b^\dagger \theta' ) |0> = 
\Psi_\mu ^\dagger \,\psi_\mu ' |0> \,,
\label{eq:D} 
\end{eqnarray}
with
\begin{equation}
\psi '   = \left(
\begin{array}{c}
\psi_1 ' \\
\psi_2 ' \\
\psi_0 '
\end{array}
\right) \equiv \left(
\begin{array}{c}
z_1 '  \\
z_2 ' \\
\theta '
\end{array}
\right) \,.
\end{equation}
(\ref{eq:D}) is exactly equal to $|\psi,1>$ in (\ref{eq:5}) if we make the 
identification
\begin{equation}
\psi_\mu ' = \frac{\psi_\mu }{|\psi|}\,.
\label{eq:ratio}
\end{equation}

For the case of general $N$ we start from the highest weight state $|N/2, N/2>$
in the $N$-fold graded symmetric tensor product $\otimes_G ^N$ of the $J_{osp(2,1)} = 1/2$
representation and the corresponding representative ${\cal U}^{\otimes_G ^N} (g)$ of $g$: 
\begin{eqnarray}
|N/2, N/2> &:=& |1/2, 1/2> \otimes_G \cdots \cdots \otimes_G 
|1/2, 1/2> \,, \nonumber \\
{\cal U}^{\otimes_G ^N} \,(g) &:=& {\cal U} \,(g) \otimes_G \cdots \cdots 
\otimes_G \,{\cal U} \,(g) \,.
\label{eq:tensorp}
\end{eqnarray}
Note that, since ${\cal U} \,(g)$ is an element of $OSp(2,1)$, 
it is even. The corresponding coherent state is
\begin{equation}
|g; N/2> = {\cal U}^{\otimes_G ^N}|N/2, N/2> = {\cal U} \,(g) \,|1/2,1/2> \otimes_G \cdots \cdots 
\otimes_G \, {\cal U} \,(g) \,|1/2,1/2> \,.
\end{equation}
Upon using (\ref{eq:D}) and (\ref{eq:ratio}), this becomes 
equivalent to (\ref{eq:5}).

\section{ On the Superspheres}

\subsection{The Commutative Supersphere $S^{(2,2)}$}

There exists an important map from the supercoherent states to 
orbits in the adjoint representation of $OSp(2,1)$ and $OSp(2,2)$. 
The former is the supersphere \cite{marmo, landi, berezin}
\begin{equation}
S^{(2,2)} = OSp(2,1)/U(1)
\end{equation}
while the latter is a closely related orbit.

We first observe that the $osp(2,1)$ generators $\Lambda_a$
($a \leq 5$) in the super Fock space are represented by
\begin{equation}
\lambda_a = \Psi_\mu ^\dagger (\Lambda_a ^{(\frac{1}{2})})_{\mu \nu} \Psi_\nu \,.
\label{eq:lLambda}
\end{equation}

The supergroup action preserves $|\psi|^2$. So consider the following map $\pi$
from the $(3,2)$-dimensional supersphere $S^{(3,2)} = \,<\psi \,: \, |\psi|^2 = {\mbox{constant}} \neq 0>$
to functions on $S^{(2,2)}$:
\begin{equation}
\pi: \quad \psi \longrightarrow \,<\psi,1|\lambda_a |\psi,1> \,:=\,
\phi_a(\psi,\bar{\psi}) \,,
\label{eq:map1}
\end{equation}
\begin{equation}
\phi_a(\psi,\bar{\psi}) = \frac{1}{|\psi|^2} \bar{\psi} 
\Lambda_a ^{(\frac{1}{2})}  \psi\,.
\label{eq:map2}
\end{equation} 
The overall phase of $\psi$ cancels out while no other degree of 
freedom is lost on R.H.S. For this reason, this is a map from $S^{(3,2)}$
to the $(2,2)$ dimensional space \footnote{In what follows we do not show the
$\bar{\psi}$ depedence of $\phi_a$ to abbreviate the notation a little bit.} 
\begin{equation}
S^{(2,2)} \,: = \, S^{(3,2)}/ \,U(1) \,= \{\phi(\psi) = 
(\phi_1(\psi), \cdots, \,\phi_5(\psi))\} \,.
\end{equation}  
$\pi$ is thus the ``super Hopf Fibration'' over $S^{(2,2)}$ 
\cite{marmo, landi}, the ``super'' generalization of the Hopf 
fibration $U(1) \, \rightarrow S^3 \,\rightarrow S^2$. 
$S^{(2,2)}$ can be thought as the supersphere generalizing $S^2$.

We now characterize $S^{(2,2)}$ as an adjoint orbit of $OSp(2,1)$. 
First observe that $\phi(\psi)$ is a (super)vector in the adjoint 
representation of $OSp(2,1)$. Under the action
\begin{equation}
\phi \,\rightarrow g\phi, \quad (g\phi)(\psi) = \phi(g^{-1}\psi), 
\quad g \in OSp(2,1) \,,
\end{equation}
it transforms by the adjoint representation $g \rightarrow \, 
Ad g$:
\begin{equation}
\phi_a(g^{-1} \psi) = 
\phi_b \,(\psi) \,(Ad \,g)_{ba} \,.
\end{equation}       

The generators of $osp(2,1)$ in the adjoint representation are
$ad \Lambda_a$ where
\begin{equation}
(ad \,\Lambda_a)_{cb} = i f_{abc} \,.
\end{equation}
From this and the infinitesimal
variations $\delta \phi(\psi) = \varepsilon_a\, ad \, \Lambda_a \,
\phi(\psi)$ of $\phi(\psi)$
under the adjoint action with $\varepsilon_i$ even and $\varepsilon_\alpha$  odd
Grassmann variables, we can verify that 
\begin{equation}
\delta ( \phi_i (\psi)^2 
+ C_{\alpha \beta} \phi_\alpha (\psi) \phi_\beta(\psi)) = 0 \,.
\end{equation}
 
Hence, $S^{(2,2)}$ is an $OSp(2,1)$ orbit with
the invariant
\begin{equation}
\phi_i(\psi)^2 + C_{\alpha \beta} \phi_\alpha (\psi) \phi_\beta(\psi) \,.
\label{eq:invariant1}
\end{equation}  
The value of the invariant can of course be changed by scaling. 
Now the even components of $\phi_a(\psi)$
are real while its odd entries depend on both $\theta$ and 
$\bar{\theta}$:
\begin{equation}
\phi_i(\psi) = \frac{1}{2} \frac{1}{|\psi|^2} \bar{z}\sigma_i z
\,,\quad \phi_4(\psi) = - \frac{1}{2} \frac{1}{|\psi|^2}
(\bar{z}_1 \theta + z_2 \bar{\theta}) \,,\quad 
\phi_5(\psi) = \frac{1}{2} \frac{1}{|\psi|^2}(- \bar{z}_2 \theta + z_1 \bar{\theta}) \,.
\end{equation}   

Let us define $\ddagger$ to be the usual adjoint operation $\dagger$
on $\Psi_\mu$ and $\Psi_\mu ^\dagger$. Then by the requirements that
\begin{eqnarray}
<\psi,1|\lambda_a |\psi,1> ^\ddagger \, &=& <\psi,1|
\lambda_a ^\ddagger |\psi,1> \,, \nonumber \\
\lambda_i ^\ddagger = \lambda_i \, &,& \lambda_\alpha ^\ddagger = C_{\alpha \beta} \lambda_\beta \,, 
\end{eqnarray}
one deduces that
\begin{equation}
\phi_i(\psi)^\ddagger = \phi_i(\psi) \quad \quad \quad
\phi_\alpha(\psi)^\ddagger = C_{\alpha \beta} \,\phi_\beta (\psi) \,,
\label{eq:real1}
\end{equation}
and that
\begin{equation}
z_i ^\ddagger = \bar{z}_i \,, \quad
\bar{z}_i ^\ddagger = z_i \,, \quad \quad  \theta^\ddagger = -\, \bar{\theta} \,,\quad
\bar{\theta}^\ddagger = \theta \,.
\label{eq:real2}
\end{equation}
(\ref{eq:real2}) preserves the condition $|\psi|^2 ={\mbox {constant}}$ and the $OSp(2,1)$ orbit.
The reality condition (\ref{eq:real1}) reduces the degrees of freedom in 
$\phi_\alpha (\psi)$ to two. The $(3,2)$ variables $\phi_a (\psi)$ 
are further reduced to $(2,2)$ on fixing the value of the invariant 
(\ref{eq:invariant1}). As $(2,2)$ is the dimension of $S^{(2,2)}$, 
there remains no further invariant in this orbit. Thus
\begin{equation}
S^{(2,2)} = < \eta^{(+)} \in {\mathbb C}^{(3,2)}: \,(\eta_i^{(+)})^2 + 
C_{\alpha \beta} \, \eta_\alpha ^{(+)} \, \eta_\beta ^{(+)} = 
\frac{1}{4}\,, (\eta_i ^{(+)})^\ddagger = \eta_i ^{(+)} \,,
(\eta_\alpha ^{(+)})^\ddagger = C_{\alpha \beta} 
\eta_\beta ^{(+)} >
\end{equation}
where we chose $\frac{1}{4}$ for the value of the invariant.

As $OSp(2,2)$ acts on $\psi$ , that is on $S^{(3,2)}$, preserving the
$U(1)$ fibres in the map $S^{(3,2)} \rightarrow S^{(2,2)}$, it has an action
on the latter. It is not the adjoint action, but closely related to it, 
as we now explain.

The nature of the $OSp(2,2)$ action on $S^{(2,2)}$ has elements of subtlety.
If $g \in OSp(2,2)$ and $\psi \in S^{(3,2)}$ then
$g \,\psi \in S^{(3,2)}$ and hence $\phi(g \,\psi) \in S^{(2,2)}$ :
\begin{equation}
\phi_i(g \,\psi)^2 + C_{\alpha \beta} \, \phi_\alpha (g \,\psi) \,
\phi_\beta(g \,\psi) = \frac{1}{4} \,, 
\end{equation}
\begin{equation}
\phi_\alpha ^\ddagger (g \,\psi) = C_{\alpha \beta} \,\phi_\beta (g \,\psi) \,.
\end{equation}
But the expansion of $\phi_\alpha (g\,\psi)$ for infinitesimal $g$ contains not only 
the even Majorana spinors
$\eta_\alpha ^{(+)}$, but also the odd ones $ \eta_\alpha ^{(-)}$, 
where $(\eta_\alpha ^{(-)}) ^\ddagger = \sum_{\beta =6,7} \tilde{C}_{\alpha \beta} \,
\eta_\beta ^{(-)}$ $(\alpha = 6,7)$. We cannot thus think of the $OSp(2,2)$ action 
as an adjoint action on the adjoint
space of $OSp(2,1)$. The reason of course is that the super Lie 
algebra  $osp(2,1)$ is not invariant under graded commutation with 
the generators $\Lambda_{6,7,8}$ of $osp(2,2)$.

Now consider the generalization of the map (\ref{eq:map1}) to 
the $osp(2,2)$ Lie algebra, 
\begin{equation}
\psi \rightarrow \,<\psi,1|\lambda_a |\psi,1> \,:=\,
\Phi_a(\psi) \,, \quad \quad a = \,1, \ldots ,8
\end{equation}
where $\lambda_a$ is given by the formula (\ref{eq:lLambda}) for all $a$
and the  $\bar{\psi}$ dependence of $\Phi_a$ has not been shown.
Just as for $OSp(2,1)$, we find,
\begin{equation}
\Phi_a(g^{-1} \psi) = \Phi_b(\psi) (Ad \,g)_{ba} \,, 
\quad \quad a,b = \,1,\ldots ,8 \,, \quad \quad g \in OSp(2,2)\,. 
\end{equation}
Thus this extended vector $\Phi(\psi) = (\Phi_1(\psi) \,,\Phi_2(\psi), 
\ldots ,\Phi_8(\psi))$ transforms as an adjoint (super)vector of 
$osp(2,2)$ under $OSp(2,2)$ action. The formula given in (\ref{eq:map2}) 
extends to this case where index $a$ there also takes the values 
$(6,7,8)$. Note that with
\begin{eqnarray}
&& \Phi_6(\psi) = \frac{1}{2} \frac{1}{|\psi|^2} (\bar{z}_1 \theta - z_2 \bar{\theta})
\,,\quad \Phi_7(\psi) = \frac{1}{2} \frac{1}{|\psi|^2}
(\bar{z}_2 \theta + z_1 \bar{\theta}) \,, \nonumber \\
&& \Phi_8(\psi) = \frac{1}{|\psi|^2}(\bar{z}_i z_i + 2 \bar{\theta} \theta) = 
(2 - \frac{1}{|\psi|^2} \bar{z}_i z_i) \,,
\end{eqnarray} 
we have
\begin{equation}
\Phi_8(\psi)^\ddagger = \Phi_8(\psi) \quad \quad
\Phi_\alpha(\psi)^\ddagger = \sum_{\beta = 6,7} \tilde{C}_{\alpha \beta} \,\Phi_\beta 
(\psi) \,,\quad \quad \alpha = 6,7
\end{equation}
showing that the new spinor $\Phi_\alpha (\psi)$, $(\alpha = 6,7)$ 
is an odd Majorana spinor as previous remarks suggested.

As $\Phi(\psi)$ transforms as an adjoint vector under $OSp(2,2)$,
the $OSp(2,2)$ Casimir function evaluated at $\Phi(\psi)$ 
is a constant on this orbit:
\begin{equation}
\Phi_i^2 (\psi) + \tilde{C}_{\alpha \beta} \,\Phi_\alpha (\psi) 
\Phi_\beta(\psi) - \frac{1}{4} \Phi_8 ^2 (\psi) = {\mbox{constant}} \,. 
\end{equation}
But we saw that the sum of the first term, and the second term 
with $\alpha \,,\beta = 4,5$ only, is invariant under $OSp(2,2)$.
Hence so are the remaining terms:
\begin{equation}
\sum_{\alpha, \beta = 6,7} \tilde{C}_{\alpha \beta} \Phi_\alpha (\psi) \Phi_\beta(\psi) -
\frac{1}{4} \Phi_8(\psi)^2 = {\mbox{constant}} \,.
\end{equation}
Its value is $- \frac{1}{4}$ as can be calculated by setting $\psi=(1,0,0)$.

In fact since the $OSp(2,1)$ orbit has the dimension of 
$S^{(3,2)}/U(1)$ and $\Phi_a(\psi) = \Phi_a(\psi \,e^{i \gamma})$ are functions of this orbit, 
we can completely express the latter in terms of $\phi(\psi)$. We find\footnote{$\Phi_{6,7,8}$
become  $-\Phi_{6,7,8}$ for $\hat{J}_{osp(2,2) -}$.} 
\begin{eqnarray} 
\Phi_6(\psi) &=& -2 (\phi_3(\psi) \phi_4(\psi) + (\phi_1(\psi) + i \phi_2(\psi)) \phi_5(\psi))
\,,\nonumber \\ 
\Phi_7(\psi) &=& 2 (\phi_3(\psi) \phi_5(\psi) - (\phi_1(\psi) -i \phi_2(\psi)) \phi_4(\psi)) \,,\nonumber \\
\Phi_8(\psi) & = & 2 (1 - \sqrt{\phi_i(\psi)^2}) \,.  
\end{eqnarray}

The generalization of the above arguments will consider the $N$-particle sector
and the map
\begin{equation}
\psi \rightarrow <\psi, N|\,\lambda_a \,|\psi,N> \,,
\end{equation}
but that brings in nothing new, R.H.S. being $N \Phi_a(\psi)$.

\subsection{The Noncommutative (Fuzzy) Supersphere $S_F^{(2,2)}$}

The fuzzy supersphere $S_F^{(2,2)}$ is obtained by replacing $\Lambda_a$ 
with the scaled $osp(2,1)$ generators
$\Lambda_a(J) = \frac{\Lambda_a}{\sqrt 2 \sqrt{J (J + \frac{1}{2})}}$ 
$(a \leq 5)$. The relations that describe $S_F^{(2,2)}$ are then
\begin{equation}
\Lambda_i(J) \Lambda_i(J) + C_{\alpha \beta} \,\Lambda_\alpha(J)
\Lambda_\beta(J) = \frac{1}{2} \,,
\label{eq:kazimir}
\end{equation}
\begin{equation} 
\lbrack \Lambda_a(J), \Lambda_b(J) \} = \frac{1}{\sqrt 2 \sqrt{J(J + 
\frac{1}{2})}}
i f_{abc} \Lambda_c(J) \,,
\label{eq:fuzzygc1}
\end{equation}
the left-hand side of (\ref{eq:kazimir}) being proportional to the Casimir operator $K_2$.
The first of these two equations describes a supersphere of radius $\frac{1}{\sqrt 2}$,
parametrized by the matrices $\Lambda_a(J)$
whereas the second one measures the amount of noncommutativity
between any two $\Lambda_a(J)$.

The graded commutative limit is recovered when $J \to \infty$,  
$\lbrack \Lambda_a(\infty), \Lambda_b(\infty) \} \to 0$.
The graded commutator in (\ref{eq:fuzzygc1}) 
naturally extends to the $osp(2,2)$ algebra $(a \leq 8)$, making the
supersymmetry richer and allowing us to compute
$\star$-products on $S_F^{(2,2)}$ as we now show.

\section{The $\star$-Products}

\subsection{$\star$-Product on $S_F ^{(2,2)}$}

First we remark that an operator (under suitable conditions) 
is completely determined by its diagonal matrix  
elements between standard coherent states \cite{Klauder}.
This is also true for diagonal matrix elements between the supercoherent states (\ref{eq:5}). 
The diagonal elements of an operator for the supercoherent states
(\ref{eq:5}) defines a function on $S^{(2,2)} _F$. We can define a $\star$-product
on functions using this map from operators to functions as we shall see below.

We first determine the above map from operators to functions. 
It is sufficient to compute the matrix elements of the coordinate operators $\lambda_a$,
the generalization to arbitrary operators 
can then be made easily. One can write $\lambda_a$ 
as in (\ref{eq:lLambda}). Proceding straightforwardly, 
the diagonal coherent state matrix element for $\lambda_a$, 
\begin{eqnarray}
W_a \,(\psi,\bar{\psi},N) &=& W_a (z_1, z_2, \bar{z}_1 , \bar{z}_2, 
\theta, \bar{\theta} ) \nonumber \\
&=& < \psi, N| \lambda_a| \psi, N>
\end{eqnarray}
can be calculated to be
\begin{equation}
W_a \,(\psi,\bar{\psi},N) = \frac{1}{|\psi|^2}\,N \bar{\psi}
\Lambda_a ^{(\frac{1}{2})} \psi\,.
\label{eq:w1}
\end{equation}
To remove the $N$ dependence in (\ref{eq:w1}), we let
\begin{equation}
{\cal W}_a \, (\psi ', \bar{\psi} ') = \frac{1}{N} W_a \,
(\psi',\bar{\psi'},N) = \bar{\psi'} \,\Lambda_a ^{(\frac{1}{2})} \,\psi' \,
\end{equation}
where $\psi ' = \frac{\psi}{|\psi|}$ and ${\cal W}_a$ is a 
superfunction of $\psi ', \bar{\psi} '$.  

We are now ready to define and compute the star product of two 
functions of the form ${\cal W}_a$ and ${\cal W}_b$. 
It depends on $N$, so we denote it by $\star_N$. It is
given by \cite{chiral, BDBJ}
\begin{equation}
{\cal W}_a \star_N {\cal W}_b \,(\psi ',\bar{\psi} ',N) 
= \frac{1}{N^2} <\psi ', N| \lambda_a \,\lambda_b | \psi ', N>
\label{eq:starproduct}
\end{equation}
which becomes, after a little manipulation
\begin{equation}
{\cal W}_a \star_N {\cal W}_b \,(\psi ',
\bar{\psi} ',N) = \frac{1}{N}
\,\bar{\psi'}\,(\Lambda_a ^{(\frac{1}{2})}  \,\Lambda_b ^{(\frac{1}{2})} )\,\psi '
+ \frac{N-1}{N}\, (\bar{\psi} '\,\Lambda_a ^{(\frac{1}{2})} \,\psi ')
(\bar{\psi} ' \,\Lambda_b ^{(\frac{1}{2})} \,\psi ') \,.
\label{eq:manip}
\end{equation}
Furthermore, since $\psi ' \Lambda_a \Lambda_b \psi'$ is ${\cal W}_a \star_1 {\cal W}_b$, 
this can be rewritten as
\begin{equation}
{\cal W}_a \star_N {\cal W}_b \,(\psi ',\bar{\psi} ',N) = \frac{1}{N}
\,{\cal W}_a \star_1 {\cal W}_b \,(\psi ', \bar{\psi} ', 1) + \frac{N-1}{N}\,
{\cal W}_a \, (\psi ', \bar{\psi} ') \,{\cal W}_b \, (\psi ', \bar{\psi} ') \,.
\label{eq:starproduct1}
\end{equation}
Introducing the matrix $K$ with
\begin{equation}
K_{ab}:= {\cal W}_a \star_1 {\cal W}_b - {\cal W}_a \, {\cal W}_b\,,
\label{eq:Kab}
\end{equation}
we can express (\ref{eq:starproduct1}) as
\begin{equation}
{\cal W}_a \star_N \,{\cal W}_b = \frac{1}{N}
K_{ab} + {\cal W}_a\,{\cal W}_b\,.
\label{eq:simpleproduct}
\end{equation} 
In this form it is apparent that in the graded commutative limit $N \to
\infty$, we recover the graded commutative product of functions 
${\cal W}_a$ and ${\cal W}_b$.

To construct the $\star$-product of arbitrary functions on 
$S^{(2,2)} _F$ we 
proceed as follows \cite{BDBJ}.
Consider first generic operators $F$ and $G$ in the representation $\hat{(\frac{N}{2})}_{osp(2,2) + }$.
We expand them in the form
\begin{eqnarray}
F &=& F^{a_1 a_2 \cdots a_N}
\, \lambda_{a_1} \otimes_G \cdots \otimes_G  \lambda_{a_N} \,, \nonumber \\
G &=& G^{b_1 b_2 \cdots b_N}
\, \lambda_{b_1} \otimes_G \cdots \otimes_G  \lambda_{b_N}
\end{eqnarray}
where  $F^{a_1 \cdots a_i a_j \cdots a_N} = (-1)^ {|a_i| |a_j|} 
F^{a_1 \cdots a_j a_i \cdots a_N}$, $|a_i| \,(mod \,2)$ being the degree of
the index $a_i$.    
The corresponding functions on $S^{(2,2)} _F$ can be read from
\begin{equation}
<\psi \,', 1| \otimes_G \cdots \otimes_G \, <\psi \,',1| \,
\left\{
\begin{array}{ll}
F^{a_1 a_2 \cdots a_N}
\, \lambda_{a_1} \otimes_G \cdots \otimes_G  \lambda_{a_N} \\
G^{b_1 b_2 \cdots b_N}
\, \lambda_{b_1} \otimes_G \cdots \otimes_G  \lambda_{b_N} \ 
\end{array}
\right.
|\psi \,',1> \otimes_G \cdots \otimes_G |\psi \,', 1>
\label{eq:tensor2}
\end{equation}
to be
\begin{eqnarray}
{\cal F}_N ({\cal W}) &=& F^{a_1 a_2 \cdots a_N}
\, {\cal W}_{a_1}\cdots  {\cal W}_{a_N} \,, \nonumber \\
{\cal G}_N ({\cal W}) &=& G^{b_1 b_2 \cdots b_N}
\, {\cal W}_{b_1}\cdots  {\cal W}_{b_N} \,.
\label{eq:tensor3}
\end{eqnarray}
In passing from(\ref{eq:tensor2}) to (\ref{eq:tensor3}), 
we have used the fact that $|\psi \,',1>$ is even. 

The $\star$-product of these functions becomes
\begin{eqnarray}
&&{\cal F}_N \star_N {\cal G}_N ({\cal W}) = (-1)^{\sum_{j>i} |a_j| |b_i|} \, 
F^{a_1 a_2 \cdots a_N} \,
({\cal W}_{a_1} \star_1 {\cal W}_{b_1}) \cdots ({\cal W}_{a_N} 
\star_1 {\cal W}_{b_N}) \, G^{b_1 b_2 \cdots b_N} \, \nonumber \\
&&= (-1)^{\sum_{j>i} |a_j| |b_i|} \, 
F^{a_1 a_2 \cdots a_N} \,
({\cal W}_{a_1} {\cal W}_{b_1} + K_{a_1b_1}) \cdots ({\cal W}_{a_N} {\cal W}_{b_N} + K_{a_Nb_N})
\,G^{b_1 b_2 \cdots b_N} \,.
\label{eq:conclusion3}
\end{eqnarray}
where we have used the formula degree of
${\cal W}_a \equiv deg {\cal W}_a = |a| \,(mod \,2)$.

In order to give the $\star$-product in (\ref{eq:conclusion3}) its final form, we proceed as follows.
First notice the identity
\begin{eqnarray}
{\cal W}_a {\cal W}_b + K_{ab} &=& {\cal W}_a (1 + \partial^{^{\!\!\!\!\!\leftarrow}}_{{\cal W}_c} 
K_{cd} \,\partial^{^{\!\!\!\!\!\rightarrow}}_{{\cal W}_d}){\cal W}_b \nonumber \\  
& \equiv & {\cal W}_a (1 + \partial^{^{\!\!\!\!\!\leftarrow}} 
K \partial^{^{\!\!\!\!\!\rightarrow}}){\cal W}_b \,. 
\end{eqnarray}
Inserting this, R.H.S. of (\ref{eq:conclusion3}) becomes
\begin{eqnarray}
&&(-1)^{\sum_{j>i} |a_j| |b_i|} \, F^{a_1 a_2 \cdots a_N} \,
({\cal W}_{a_1}(1 + \partial^{^{\!\!\!\!\!\leftarrow}} 
K \partial^{^{\!\!\!\!\!\rightarrow}}){\cal W}_{b_1}) \cdots ({\cal W}_{a_N}
(1 + \partial^{^{\!\!\!\!\!\leftarrow}} \, 
K \, \partial^{^{\!\!\!\!\!\rightarrow}}) {\cal W}_{b_N})
\,G^{b_1 b_2 \cdots b_N} \nonumber \\
&& \equiv  {\cal F}_N ({\cal W}) (1 + \partial^{^{\!\!\!\!\!\leftarrow}} 
K \partial^{^{\!\!\!\!\!\rightarrow}})_{11} \cdots (1 + \partial^{^{\!\!\!\!\!\leftarrow}} 
K \partial^{^{\!\!\!\!\!\rightarrow}})_{ii} \cdots
(1 + \partial^{^{\!\!\!\!\!\leftarrow}} 
K \partial^{^{\!\!\!\!\!\rightarrow}})_{NN} \,{\cal G}_N ({\cal W}) \,, 
\label{eq:aux}
\end{eqnarray}
where in the second line of (\ref{eq:aux}) we have introduced an auxilary notation
whose meaning will now be explained. First we note that we have written
$(1 + \partial^{^{\!\!\!\!\!\leftarrow}} K \partial^{^{\!\!\!\!\!\rightarrow}})_{ii}$
for $1 + (\partial^{^{\!\!\!\!\!\leftarrow}} K \partial^{^{\!\!\!\!\!\rightarrow}})_{ii}$, purely for
notational convenience. Now the meaning of the subscripts in
$(\partial^{^{\!\!\!\!\!\leftarrow}} K \partial^{^{\!\!\!\!\!\rightarrow}})_{ii}$ is as follows:
Consider the expansion of ${\cal F}_N$ and ${\cal G}_N$ given in (\ref{eq:tensor3}). There are 
$N$ slots in both ${\cal F}_N$ and ${\cal G}_N$ each occupied by some ${\cal W}_a$ and ${\cal W}_b$
respectively. The first subscript $i$ means that $\partial^{^{\!\!\!\!\!\leftarrow}}$ in
$(\partial^{^{\!\!\!\!\!\leftarrow}} K \partial^{^{\!\!\!\!\!\rightarrow}})_{ii}$ acts only on the
${\cal W}_a$ in the $i^{th}$ slot (counting from left) in the expansion of ${\cal F}_N$ in
(\ref{eq:tensor3}), whereas the second subscript $i$ means that
$\partial^{^{\!\!\!\!\!\rightarrow}}$ in $(\partial^{^{\!\!\!\!\!\leftarrow}} K
\partial^{^{\!\!\!\!\!\rightarrow}})_{ii}$ acts only on the ${\cal W}_b$ in the $i^{th}$
slot(counting from left) in the expansion of ${\cal G}_N$ in (\ref{eq:tensor3}). We also note the
following:
\begin{eqnarray}
&& {\it i)} \,\,{\cal F}_N ({\cal W}) \,(\partial^{^{\!\!\!\!\!\leftarrow}} 
K \partial^{^{\!\!\!\!\!\rightarrow}})_{ii} \,{\cal G}_N ({\cal W}) = {\cal F}_N ({\cal W}) \,
(\partial^{^{\!\!\!\!\!\leftarrow}} K \partial^{^{\!\!\!\!\!\rightarrow}})_{jj} \,{\cal G}_N ({\cal W})
\,, \nonumber \\
&& {\it ii)} \,\,{\cal F}_N ({\cal W}) ({\partial^{^{\!\!\!\!\!\leftarrow}}}_{{\cal W}_{a_j}})_i \,
({\partial^{^{\!\!\!\!\!\leftarrow}}}_{{\cal W}_{a_k}})_i
= (\partial^{^{\!\!\!\!\!\rightarrow}}_{{\cal W}_{b_j}})_i \,
(\partial^{^{\!\!\!\!\!\rightarrow}}_{{\cal W}_{b_k}})_i \,
{\cal G}_N ({\cal W}) = 0 \,.  
\end{eqnarray}   
Here ${\cal F}_N ({\cal W}) ({\partial^{^{\!\!\!\!\!\leftarrow}}}_{{\cal W}_{a_m}})_i \,
((\partial^{^{\!\!\!\!\!\rightarrow}}_{{\cal W}_{b_m}})_i \,{\cal G}_N ({\cal W}))$
means that we apply the derivative ${\partial^{^{\!\!\!\!\!\leftarrow}}}_{{\cal W}_{a_m}} \,
(\partial^{^{\!\!\!\!\!\rightarrow}}_{{\cal W}_{b_m}})$ only on the $i^{th}$ slot in the expansion
of ${\cal F}_N ({\cal W}) ({\cal G}_N ({\cal W}))$ in the sense explained above.
It takes a simple relabeling of the indices and the use of symmetries of ${\cal F}_N ({\cal W})$  
and ${\cal G}_N ({\cal W})$ to prove {\it i}) while {\it ii}) is obvious by inspection.

We can write (\ref{eq:aux}) as 
\begin{eqnarray}
&&{\cal F}_N \star_N {\cal G}_N ({\cal W}) = {\cal F}_N \,{\cal G}_N ({\cal W}) \nonumber \\ 
\quad  &&+\sum_{m = 1}^N \frac{N!}{(N-m)!m!} \,{\cal F}_N  ({\cal W}) \,((\partial^{^{\!\!\!\!\!\leftarrow}}
\,K \, \partial^{^{\!\!\!\!\rightarrow}})_{11} \cdots
(\partial^{^{\!\!\!\!\!\leftarrow}} \,K \,\partial^{^{\!\!\!\!\rightarrow}})_{ii} \cdots 
(\partial^{^{\!\!\!\!\!\leftarrow}} \,K \,\partial^{^{\!\!\!\!\rightarrow}})_{mm}) \,{\cal G}_N ({\cal W}) \,.
\label{eq:binomial1}
\end{eqnarray}

Now observe the following identities: 
\begin{eqnarray}
{\cal F}_N({\cal W}) \,(\partial^{^{\!\!\!\!\!\leftarrow}}_{{\cal W}_{a_1}})_{1} \cdots
(\partial^{^{\!\!\!\!\!\leftarrow}}_{{\cal W}_{a_1}})_{i}
\cdots (\partial^{^{\!\!\!\!\!\leftarrow}}_{{\cal W}_{a_m}})_{m} 
&=& \frac{(N-m)!}{N!} \,{\cal F}_N({\cal W}) \,\partial^{^{\!\!\!\!\!\leftarrow}}_{{\cal W}_{a_1}}
\cdots \partial^{^{\!\!\!\!\!\leftarrow}}_{{\cal W}_{a_i}} 
\cdots \partial^{^{\!\!\!\!\!\leftarrow}}_{{\cal W}_{a_m}} \,, \nonumber \\
(\partial^{^{\!\!\!\!\!\rightarrow}}_{{\cal W}_{b_1}})_{1}
\cdots (\partial^{^{\!\!\!\!\!\rightarrow}}_{{\cal W}_{b_m}})_{i}
\cdots (\partial^{^{\!\!\!\!\!\rightarrow}}_{{\cal W}_{b_m}})_{m} \,{\cal G}_N({\cal W})
&=& \frac{(N-m)!}{N!} \,\partial^{^{\!\!\!\!\!\rightarrow}}_{{\cal W}_{b_1}}
\cdots \partial^{^{\!\!\!\!\!\rightarrow}}_{{\cal W}_{b_i}}
\cdots \partial^{^{\!\!\!\!\!\rightarrow}}_{{\cal W}_{b_m}} {\cal G}_N({\cal W}) \,.
\label{eq:der1}
\end{eqnarray}

To  facilitate the use of (\ref{eq:der1}) in (\ref{eq:binomial1}) we define the ordering $\vdots \cdots \vdots$
in which $\partial^{^{\!\!\!\!\!\leftarrow}}_{{\cal W}_{a_i}}$
($\partial^{^{\!\!\!\!\!\rightarrow}}_{{\cal W}_{b_i}}$) are moved to the left (right) extreme and    
they act on everything to their left (right) :
\begin{eqnarray}
&&\vdots \,(1 + \partial^{^{\!\!\!\!\!\leftarrow}}\,
K \,\partial^{^{\!\!\!\!\rightarrow}})^N \,\vdots \, =
1 + \sum_{m = 1}^N \frac{N!}{(N-m)!m!} \,\vdots \,\underbrace{(\partial^{^{\!\!\!\!\!\leftarrow}}
\,K \,\partial^{^{\!\!\!\!\rightarrow}})
\cdots (\partial^{^{\!\!\!\!\!\leftarrow}}
\,K \,\partial^{^{\!\!\!\!\rightarrow}})}_{m \,factors} \,\vdots \, \nonumber \\
&& \equiv 1 + \sum_{m = 1}^N \frac{N!}{(N-m)! \,m!} \,(-1)^\Delta \,
\partial^{^{\!\!\!\!\!\leftarrow}}_{{\cal W}_{a_1}}
\cdots \partial^{^{\!\!\!\!\!\leftarrow}}_{{\cal W}_{a_m}} \,K_{a_1 b_1} \cdots K_{a_m b_m} \,  
\partial^{^{\!\!\!\!\rightarrow}}_{{\cal W}_{b_1}} \cdots
\partial^{^{\!\!\!\!\rightarrow}}_{{\cal W}_{b_m}} \,.
\label{eq:ordering1}
\end{eqnarray}
where $\Delta =  \sum_{j>i}^m (|a_j| |a_i| + |a_j| |b_i| + |b_j| |b_i|) \, (mod \,2)$ is the overall
degree due to moving the derivatives to their positions in (\ref{eq:ordering1}). 
Using (\ref{eq:der1}) and (\ref{eq:ordering1}) in (\ref{eq:binomial1}) we write our
$\star$-product in its final form
\begin{equation}
{\cal F}_N \star_N {\cal G}_N ({\cal W}) = {\cal F}_N \,{\cal G}_N ({\cal W}) +
\sum_{m = 1}^N \frac{(N-m)!}{N! \,m!} \,{\cal F}_N  ({\cal W}) \,\,\vdots
\underbrace{(\partial^{^{\!\!\!\!\!\leftarrow}}
\,K \,\partial^{^{\!\!\!\!\rightarrow}}) \cdots (\partial^{^{\!\!\!\!\!\leftarrow}}
\,K \,\partial^{^{\!\!\!\!\rightarrow}})}_{m \,factors}  
\,\vdots \, {\cal G}_N ({\cal W}) \,.
\label{eq:finalstar}
\end{equation}  
It depends on the ordering introduced in (\ref{eq:ordering1}) and not on the auxilary notation
of (\ref{eq:aux}) and (\ref{eq:der1}). From this formula it is apparent that, in the graded commutative limit
$(N \to \infty)$, we get back the ordinary pointwise multiplication ${\cal F}_N \,{\cal G}_N ({\cal W})$.

\subsection{$\star$-Product on Fuzzy ``Sections of Bundles''}

In this subsection we extend the $\star$-product found above
to functions obtained from matrix 
elements of annihilation-creation operators between 
vectors of different $N$, using the ideas of \cite{private}.
Linear operators between vector spaces of different $N$
correspond to sections of bundles \cite{GKP2, chiral, monopole} so that in this manner, we extend our
supersymmetric $\star$-product to sections of bundles. 
The results below also
provide an alternative way to compute the $\star$-products
in (\ref{eq:starproduct1}) and (\ref{eq:finalstar}).
 
First note that with $\lbrack \Psi_\mu , 
\Psi_\nu ^\dagger \} = \delta_{\mu \nu}$, we have
\begin{eqnarray}
\Psi_\mu |\psi, N> &=& \sqrt{N} \,
\frac{\psi_\mu}{|\psi|}\,|\psi, N - 1> \,, \\
< \psi, N|\Psi_\mu ^\dagger &=& < \psi, N - 1|\sqrt{N}\,
\frac{\bar{\psi}_\mu }{|\psi|}\,.
\end{eqnarray}
We now let $\psi' = \psi/|\psi|$ as before and define
\begin{eqnarray}
S_\mu  &:=& \Psi_\mu \,\frac{1}{\sqrt{N}} = \frac{1}{\sqrt{N + 1}}\,
\Psi_\mu \,, \\
S_\mu ^\dagger  &:=& \frac{1}{\sqrt{N}}\, \Psi_\mu ^\dagger = 
\Psi_\mu ^\dagger \,\frac{1}{\sqrt{N + 1}}
\end{eqnarray}
where $N = \Psi_\mu^{\dagger} \, \Psi_\mu$ is the number 
operator. Then
\begin{eqnarray}
S_\mu |\psi ', N> &=& \psi_\mu ' |\psi ', N - 1> \,, \\
<\psi ', N| S_\mu ^\dagger &=& <\psi ', N - 1| \bar{\psi_\mu} ' \,.
\label{eq:vectors}
\end{eqnarray}
Furthermore, we have that $\lbrack S_\mu , S_\nu \} = 
\lbrack S_\mu ^\dagger , S_\nu ^\dagger \} =0$ while after 
a small calculation we get
\begin{equation}
\lbrack S_\mu , S_\nu ^\dagger \} = \frac{1}{N + 1}\,
\left( \delta_{\mu \nu}- (-1)^{|S_\mu| |S_\nu|}\, S_\nu ^\dagger
S_\mu \right) \,
\label{eq:gcommutator}
\end{equation}
where $|S_\mu|$ denotes the degree of $S_\mu$. Using 
(\ref{eq:vectors}), we also get
\begin{eqnarray}
<\psi ', N - 1|S_\mu|\psi ', N> &=& \psi_\mu ' \\
<\psi ', N |S_\mu ^\dagger |\psi ' ,N - 1> &=& \bar{\psi}_\mu '\,.
\end{eqnarray}
Thus, the $\star$-product of $\psi '$ with $\bar{\psi} '$ is given by 
\begin{eqnarray}
\psi_\mu ' \star \bar{\psi}_\nu ' &=&  <\psi ', N |S_\mu 
S_\nu ^\dagger |\psi ', N> \nonumber \\
&=& <\psi ', N | (-1)^{|S_\mu| |S_\nu|}\, \frac{N}{N + 1}\,S_\nu ^\dagger
S_\mu + \frac{1}{N + 1}\,\delta_{\mu \nu} |\psi ', N> \nonumber \\
&=& \frac{N}{N + 1}\, \psi_\mu ' \bar{\psi}_\nu ' + \frac{1}{N + 1}\,
\delta_{\mu \nu} \,.
\end{eqnarray}
Here we have used (\ref{eq:gcommutator}) and the fact that 
$\psi_\mu ' \bar{\psi}_\nu ' = (-1)^{|S_\mu| |S_\nu|} \,\bar{\psi}_\nu ' 
\,\psi_\mu '$ to get rid of $(-1)^{|S_\mu| |S_\nu|}$. Rearreaging the 
last result we can write
\begin{eqnarray}
\psi_\mu ' \star \bar{\psi}_\nu ' &=& \frac{1}{N + 1}\, \Omega_{\mu \nu}
+ \psi_\mu ' \bar{\psi}_\nu ' \,, \nonumber \\
\Omega_{\mu \nu} &\equiv& 
\delta_{\mu \nu} - \psi_\mu '\, \bar{\psi}_\nu ' \,. 
\label{eq:63}
\end{eqnarray} 
As an easy check of our results, we can compute
${\cal W} _a \star_N {\cal W}_b$, using the method above. 
First note that 
\begin{equation}
{\cal W}_a = \bar{\psi} ' \,\Lambda_a \,\psi ' = \, 
<\psi ', N |S ^\dagger \, \Lambda_a \,S|\psi ', N>\,.
\end{equation}
Hence
\begin{eqnarray}
{\cal W}_a \star_N {\cal W}_b &=& 
<\psi ', N |S_\mu ^\dagger \,(\Lambda_a)_{\mu \nu} \,S_\nu
S_\alpha ^\dagger \,(\Lambda_b)_{\alpha \beta} \,S_\beta |\psi ', N>
\nonumber \\
&=& \bar{\psi}_\mu ' \,(\Lambda_a)_{\mu \nu}\, \left(
\frac{1}{N}\,\Omega_{\nu \alpha} + \psi_\nu ' \bar{\psi}_\alpha ' \right)
\,(\Lambda_b)_{\alpha \beta} \, \psi_\beta ' \nonumber \\
&=& \bar{\psi}_\mu ' \,(\Lambda_a)_{\mu \nu}\, \left(
\frac{1}{N}\,\delta_{\nu \alpha} + \frac{N - 1}{N}\,
\psi_\nu ' \bar{\psi}_\alpha ' \right) \,(\Lambda_b)_{\alpha \beta} 
\, \psi_\beta ' \nonumber \\
&=& \frac{1}{N}\, {\cal W}_a \star_1 {\cal W}_b + 
\frac{N - 1}{N}\, {\cal W}_a \,{\cal W}_b \,
\label{eq:conclusion2}
\end{eqnarray}
which is (\ref{eq:starproduct1}).

Comparing the second line of the last equation with
(\ref{eq:simpleproduct}) we get the important 
result\footnote{We thank Peter Pre\v{s}najder for a discussion
leading to (\ref{eq:Kprojector})} 
\begin{eqnarray}
K_{ab} & = & ({\cal W}_a \,\partial^{^{\!\!\!\!\!\leftarrow}}_\mu)
\,\Omega_{\mu \nu} \,(\vec{\bar{\partial}}_\nu \,{\cal W}_b) \nonumber \\
&\equiv &{\cal W}_a \,\partial^{^{\!\!\!\!\!\leftarrow}} \,\Omega \,\vec{\bar{\partial}} \,{\cal W}_b \,, 
\label{eq:Kprojector}
\end{eqnarray}
where $\partial^{^{\!\!\!\!\!\leftarrow}} \,\Omega \,\vec{\bar{\partial}} \equiv 
\partial^{^{\!\!\!\!\!\leftarrow}}_\mu \,\Omega_{\mu \nu} \,\vec{\bar{\partial}}$ and
$\partial_\mu = \frac{\partial}{\partial \,\psi_\mu '}$.

We would like to note that this result can be used to write (\ref{eq:finalstar}) in terms
of $\partial^{^{\!\!\!\!\!\leftarrow}} \,\Omega \,\vec{\bar{\partial}}$.
To this end we write (\ref{eq:conclusion3}) as 
\begin{equation}
(-1)^{\sum_{j>i} |a_j| |b_i|} \, F^{a_1 a_2 \cdots a_N} \,
({\cal W}_{a_1}(1 + \partial^{^{\!\!\!\!\!\leftarrow}}
\,\Omega \,\vec{\bar{\partial}}){\cal W}_{b_1}) \cdots ({\cal W}_{a_N}
(1 + \partial^{^{\!\!\!\!\!\leftarrow}}
\,\Omega \,\vec{\bar{\partial}})  {\cal W}_{b_N}) \,G^{b_1 b_2 \cdots b_N} \,. 
\label{eq:omegastar1}
\end{equation}
Carrying out a calculation similar to the one given in (\ref{eq:aux}) through (\ref{eq:finalstar}),
one finally finds
\begin{equation}
{\cal F}_N \star_N {\cal G}_N ({\cal W}) = {\cal F}_N \,{\cal G}_N ({\cal W}) +
\sum_{m = 1}^N \frac{(N-m)!}{N! m!} \,{\cal F}_N  ({\cal W}) \,\,\vdots
\underbrace{(\partial^{^{\!\!\!\!\!\leftarrow}}
\,\Omega \,\vec{\bar{\partial}}) \cdots (\partial^{^{\!\!\!\!\!\leftarrow}}
\,\Omega \,\vec{\bar{\partial}})}_{m factors} \, \vdots \, {\cal G}_N ({\cal W}) \,,
\label{eq:omegastar2}
\end{equation}
where now $\vdots \cdots \vdots$ takes $\partial^{^{\!\!\!\!\!\leftarrow}}$ and 
$\vec{\bar{\partial}}$ to the left and right extreme respectively.
(When $\partial^{^{\!\!\!\!\!\leftarrow}}$'s and $\vec{\bar{\partial}}$'s
are moved in this fashion, the phases coming from the graded commutators should be included
just as for (\ref{eq:finalstar})). 
  
It can be explicitly shown that $\Omega = (\Omega_{\mu\nu})$ is a projector, i.e.,
\begin{equation}
\Omega^2 = \Omega \quad \quad {\mbox{and}} \quad \quad
\Omega^\ddagger = \Omega \,.
\end{equation}
Due to (\ref{eq:Kprojector}), the last equation implies similar
properties for \footnote{Following the conventions of \cite{GKP1},
we consider all the indices down through out this paper. In what follows
the relevant object under investigation is ${\cal K}_{ab}$ corresponding to
$K_a {}^b$ in a notation where indices are raised and lowered by the metric.
To stick with the conventions of \cite{GKP1}, we write (\ref{eq:conv}) and proceed accordingly.} 
\begin{equation}
{\cal K}_{ab} \equiv (K \,S^{-1})_{ab} \,.
\label{eq:conv}
\end{equation}
which we discuss next.

\section{More on the Properties of ${\cal K}_{ab}$}

A closer look at the properties of ${\cal K}_{ab} \equiv (K \,S^{-1})_{ab}$, where
\begin{eqnarray}
K_{ab}(\psi) &=& {\cal W}_a \star_1 {\cal W}_b (\psi) - {\cal W}_a (\psi) \,{\cal W}_b (\psi) \nonumber \\ 
&=& <\psi,1|\lambda_a \lambda_b |\psi,1> - <\psi,1|\lambda_a |\psi,1>
<\psi,1|\lambda_b|\psi,1>\,,
\label{eq:conv2}
\end{eqnarray}
will give us  more insight on the structure of the 
$\star$-product found in the previous section.
First note that ${\cal K}_{ab}$ depends on both $\psi$ and $\bar{\psi}$. We denote this
dependence by ${\cal K}_{ab}(\psi)$ for short, omitting to write the
$\bar{\psi}$ dependence. Now we would like to show that
the matrix ${\cal K}(\psi) \,= ({\cal K}_{ab}(\psi))$ is a projector.

We first recall that in $(\hat{\frac{1}{2}})_{osp(2,2) + }$, representation of $osp(2,2)$
the highest and lowest weight states are given by
\begin{equation}
|J, J_3> = \left\{
\begin{array}{ll}
|1/2,\, 1/2> & = {\mbox{highest weight state}}, \\
|1/2,-1/2>  & = {\mbox{lowest weight state}}
\end{array}
\right.
\label{eq:weigths}
\end{equation}

We note that, starting from the lowest weight state
$|1/2, -1/2> = \Psi_2 ^\dagger|0>$, one could construct another supercoherent state,
expressed by a formula similar to (\ref{eq:D}).
Now consider the following fiducial points for ${\cal W}(\psi)$ at $\psi = \psi^0 = (1,0,0)$
obtained from computing ${\cal W}_a (\psi ^0)$ in the supercoherent states induced from the
states given in (\ref{eq:weigths}):  
\begin{equation}
{\cal W}^{\pm} (\psi^0) = ({\cal W}_1 (\psi^0) \cdots {\cal W}_8 (\psi^0)) = 
(0,0,\pm 1/2,0,0,0,0,1)\,.
\label{eq:fudicial}
\end{equation} 
In (\ref{eq:fudicial}) $+(-)$ correponds to upper(lower) entries in (\ref{eq:weigths}) and
the calculation is done using (\ref{eq:w1}).
 
Although not essential in what follows, we remark that
${\cal W}^- (\psi = (1,0,0)) = {\cal W}^+ (\psi = (0,1,0))$, that is,
\begin{equation}
{\cal W}_a ^- (\psi ^0) = {\cal W}_b ^+(\psi^0) \,(Ad \,e^{i \pi \Lambda_2 ^{(\frac{1}{2})}})_{ba} \,.
\end{equation}
 
Note that all other points in $S_F^{(2,2)}$ can be obtained from 
${\cal W}^{\pm} (\psi^0)$ by the adjoint action of the group, i.e., 
\begin{equation}
{\cal W}_a ^{\pm} (\psi) = {\cal W}_b ^{\pm} (\psi^0) (Ad\,g^{-1})_{ba} \,
\end{equation}
where $\psi = g \psi^0$.

We define ${\cal K}^\pm (\psi^0)$ using ${\cal W}^\pm (\psi^0)$ for ${\cal W}$,
(\ref{eq:conv}) and (\ref{eq:conv2}).
The matrices ${\cal K}^{\pm}(\psi^0)$ when computed at the fiducial points (using for instance
(\ref{eq:nontyp},\ref{eq:manip},\ref{eq:Kab})) have the block diagonal forms
\begin{equation}
{\cal K} ^\pm \,(\psi^0) =
({\cal K}_{ab} ^\pm \,(\psi^0)) = \left(
\begin{array}{ccc}
\left(\frac{1}{2}\,\delta_{ij} \pm \frac{i}{2}\,
\epsilon_{ij3} -2 \,{\cal W}_i ^\pm (\psi^0) \,({\cal W}_j ^\pm \,(\psi^0)) 
\right)_{3 \times 3} & & \\
 & \left( \Sigma^\pm _{\alpha \beta} \right)_{4 \times 4} & \\
 & & 0
\end{array}
\right) _{8 \times 8}
\label{eq:matrixK}
\end{equation}
with
\begin{equation}
\Sigma^\pm = ( \Sigma^\pm _{\alpha \beta}) =
\frac{1}{4}\left(
\begin{array}{cc}
1 \pm \sigma_3 & -(1 \pm \sigma_3) \\
-(1 \pm \sigma_3) & 1 \pm \sigma_3
\end{array}
\right)
\end{equation}
where the upper (lower) sign stands for the upper
(lower) sign in ${\cal W} ^\pm \,(\psi^0)$.
The matrices ${\cal K}^\pm \,(\psi^0)$ have only even components and do not mix the
$1,2,3,8$ and $4,5,6,7$ entries of a vector. So its grade adjoint
is its ordinary adjoint $\dagger$.
Now from (\ref{eq:matrixK}) it is straightforward to check the relations
\begin{equation}
\begin{array}{ccc}
&&({\cal K}^{\pm} \,(\psi^0))^2 = {\cal K}^{\pm} \,(\psi^0) \,,\\
&&({\cal K}^{\pm} \,(\psi^0))^\ddagger = {\cal K}^{\pm} \,(\psi^0)\,,\\
&&{\cal K}^+ \,(\psi^0) \,{\cal K}^- \,(\psi^0) = 0
\end{array}
\end{equation}
which show that ${\cal K}^{\pm} \,(\psi^0)$ are orthogonal projectors. By the adjoint 
action of the group, we have
\begin{equation}
{\cal K}_{ab} ^{\pm} \,(\psi) = ((Ad \,g)^T)^{-1}_{ad} \,{\cal K}_{de} ^{\pm} \,(\psi^0)
\,(Ad \, g)^T _{eb} \,,
\label{eq:Adj}
\end{equation}
with $T$ denoting transpose, implying that ${\cal K} ^{\pm} \,(\psi)$ are projectors
for all $g \in \,OSp(2,2)$.

We further observe that a super-analogue ${\cal J}$ 
of the complex structure can be defined over the 
supersphere. To show this, following \cite{BDBJ}, we first observe that
the projective module for ``sections of the supertangent bundle'' over
$S^{(2,2)}$ is ${\cal P}{\cal A}^8$, where ${\cal A}$ is the algebra of superfunctions over
$S^{(2,2)}$, ${\cal A}^8 = {\cal A} \otimes_{{\mathbb C}} \,{\mathbb C}^8$ and 
${\cal P}{(\psi)} = {\cal K}^+ \,(\psi) + {\cal K}^- \,(\psi)$ is a projector. 
(For details, see \cite{BDBJ}.) The super-complex structure then is given by
the matrix ${\cal J}$ with elements  
\begin{equation}
{\cal J}_{ab} (\psi) = i\,({\cal K}^+ - {\cal K}^-)_{ab} (\psi) \,.
\label{eq:complexstruc}
\end{equation}
It acts on  ${\cal P}{\cal A}^8$. Since
\begin{equation}
{\cal J}^2 \,(\psi) \Big{\vert}_{{\cal P}{\cal A}^8} = - {\cal P} (\psi) \,\Big{\vert}_{{\cal P}{\cal A}^8}
= -{\bf 1} \,\Big{\vert}_{{\cal P}{\cal A}^8}
\end{equation}
($\delta \Big{\vert}_\varepsilon$ denoting the restriction of $\delta$ to
$\varepsilon$),
it defines a super complex structure. Furthermore, due to the relation
\begin{equation}
{\cal J} \Big{\vert}_{{\cal K} ^{\pm} {\cal A}^8} = \pm i \,\Big{\vert}_{{\cal K} ^{\pm} {\cal A}^8} \,,
\end{equation}
${\cal K} ^{\pm} {\cal A}^8$ give the ``holomorphic''
and ``anti-holomorphic'' parts of ${\cal P}{\cal A}^8$.

Finally we also have
\begin{equation} 
{\cal K} ^{\pm} \,(\psi) = \frac{1}{2}\,(-{\cal J}^2 \mp i{\cal J})(\psi) \,.
\label{eq:projectorK}
\end{equation}

\section{Discussion and Conclusions}

In this article we have constructed the $\star$-product of 
functions on $S_F^{(2,2)}$. Our central result is given in 
(\ref{eq:finalstar}). A consequence of (\ref{eq:starproduct1})
is the graded commutator of the $\star$-product
\begin{equation}
\lbrack {\cal W}_a, {\cal W}_b \}_{\star_N} = \frac{i}{N} 
f_{abc} {\cal W}_c
\end{equation}
which generalizes a familiar result for the usual $\star$-products.
A special case of our result for the $\star$-product follows
if we restrict ourselves to the even subspace $S_F^2$ of $S_F^{(2,2)}$, namely the fuzzy sphere.
In this case, we get from (\ref{eq:starproduct1}) and (\ref{eq:finalstar}),
\begin{eqnarray}
&&{\cal F}_N \star_N {\cal G}_N \,({\cal W}) =
{\cal F}_N \,{\cal G}_N ({\cal W}) + \nonumber \\ 
&&\quad  \sum_{m=1}^N \frac{(N - m)!}{N! \,m!}\,
(\partial_{a_1} \cdots \partial_{a_m} {\cal F}_N \,({\cal W})) 
\,\,K_{a_1 b_1} \cdots K_{a_m b_m}\,\,
(\partial_{b_1} \cdots \partial_{b_m} {\cal G}_N \,({\cal W}))\,,    
\end{eqnarray}
$$
\partial_{a_i} \equiv \partial_{{\cal W}_{a_i}}\,, \quad \partial_{b_j} \equiv \partial_{{\cal W}_{b_j}} \,,
$$
which is the formula given in \cite{chiral, BDBJ}.

There have been developments
in writing sigma models in $S_F^2$ using Bott projectors \cite{bott}.
It appears that the projector $\Omega$ introduced in (\ref{eq:Kprojector}) 
is the supersymmetric version of Bott projector (for Chern class $1$)  and
can be the starting point for constructing  sigma models on $S_F^{(2,2)}$. 
We will develop this idea in another article. 
\vspace{.5cm}

{\bf Acknowledgements}

\vspace{.5cm}

The authors are indebted to Denjoe O'Connor and Peter Pre\v{s}najder
for useful discussions and Peter Pre\v{s}najder for detailed comments on paper.
E.R. would like to thank Department
of Physics of Syracuse University for hospitality during
his stay in Syracuse and acknowledges support from a CONACyT 
post-doctoral fellowship. The work of A.P.B and S.K. was supported 
in part by DOE and NSF under contract numbers
DE-FG02-85ER40231 and INT9908763 respectively.

\end{document}